\documentclass{sf2a-conf2024}
\usepackage{graphicx}
\usepackage[colorlinks=true, citecolor=blue, linkcolor=blue]{hyperref}
\usepackage[]{natbib}  
\usepackage{epstopdf}
\usepackage{xcolor}
\usepackage{amsmath,amssymb}

\def\BibTeX{{\rm B\kern-.05em{\sc i\kern-.025em b}\kern-.08em
    T\kern-.1667em\lower.7ex\hbox{E}\kern-.125emX}}
\bibpunct{(}{)}{;}{a}{}{,}  

\begin{document}

\TitreGlobal{SF2A 2024}


\title{The PeV Frontier: Status of Gamma-ray astronomy after two decades with H.E.S.S., MAGIC, VERITAS and the new window recently opened by HAWC and LHAASO}

\runningtitle{Status of Galactic $\gamma$-ray astronomy}

\author{J. Devin}\address{Laboratoire Univers et Particules de Montpellier, Université de Montpellier, CNRS/IN2P3, F-34095 Montpellier, France}

\setcounter{page}{237}


\maketitle


\begin{abstract}
One of the main purposes in $\gamma$-ray astronomy is linked to the origin of Galactic cosmic rays. Unlike cosmic rays, $\gamma$ rays can be used to probe their production sites in the Galaxy and to find which type of astrophysical sources is able to accelerated particles up to PeV energies. Twenty years of observations with current Imaging Atmospheric Cherenkov Telescopes (H.E.S.S., MAGIC and VERITAS) provided an unprecedented view of the very-high-energy $\gamma$-ray sky and a large variety of Galactic sources which are prominent TeV emitters, such as supernova remnants, pulsar wind nebulae, massive stellar clusters and binary systems, in addition to a large fraction of unidentified TeV sources. For a long time, supernova remnants were the most promising candidates for the main source of Galactic cosmic rays, but the new window of ultra-high-energy $\gamma$ rays recently opened by HAWC and LHAASO gave unexpected results and demonstrated the need to re-evaluate some scenarios and to revise some of our definitions. The highest-energy $\gamma$-ray sources are not associated with standard candidates for the main source of Galactic cosmic rays and challenged our usual paradigms, highlighting the vastness of what needs to be explored and understood in the next decades.
\end{abstract}



\section{Introduction}
Although cosmic rays (CRs) were detected more than one century ago, their origin is still debated. These charged particles (containing 90\% of protons, 9\% of heavier nuclei and 1\% of electrons) are accelerated to relativistic energies before travelling and reaching the Earth. CRs with energies up to $10^{15}-10^{17}$ eV are thought to be accelerated within the Milky Way because they are confined by the Galactic magnetic field. Understanding the origin of Galactic CRs requires to find which type of astrophysical sources is able to accelerate hadrons up to PeV energies, the so-called "PeVatrons". Because these charged particles are deflected by the Galactic magnetic field, we rely on the $\gamma$-ray emission they produce when being accelerated to pinpoint their production site. Hadronic CRs when interacting with ambient matter produce neutral pions that rapidly decay into $\gamma$ rays, while leptonic CRs emit $\gamma$ rays when interacting with photon fields (through Inverse Compton scattering), ambient matter (Bremsstrahlung) or magnetic field (synchrotron emission). Because $\gamma$ rays can be produced both by protons and electrons, understanding the nature (dominantly hadronic or leptonic) of the $\gamma$-ray emission from astrophysical sources is not always straightforward. A smoking gun for proton acceleration is the detection of the so-called "pion bump" near 70 MeV, indicative of proton-proton interactions, which has been already observed in some supernova remnants. For a long time, it was thought that detecting $\gamma$-ray photons with energy $E_{\gamma} \geq 100$ TeV would be evidence for hadronic emission because at these energies, leptons are in the Klein-Nishina regime and are not able to efficiently produce $\gamma$ rays through Inverse Compton scattering. For this reason, the detection of  $\gamma$-ray emission up to 100 TeV without any curvature or cutoff was considered as evidence for a PeVatron (because $E_{p} \sim 10 \times E_{\gamma}$), and this term implicitly referred to a hadronic PeVatron. After more than 20 years with current Imaging Atmospheric Cherenkov Telescopes (IACTs) and first light from current Extensive Air Shower (EAS) experiments, we will see that this last statement is no longer valid.

\section{The Glory Days of current IACTs: What did we learn?} \label{sec:IACTs}

\begin{figure}[ht!]
 \centering
 \includegraphics[width=1\textwidth,clip]{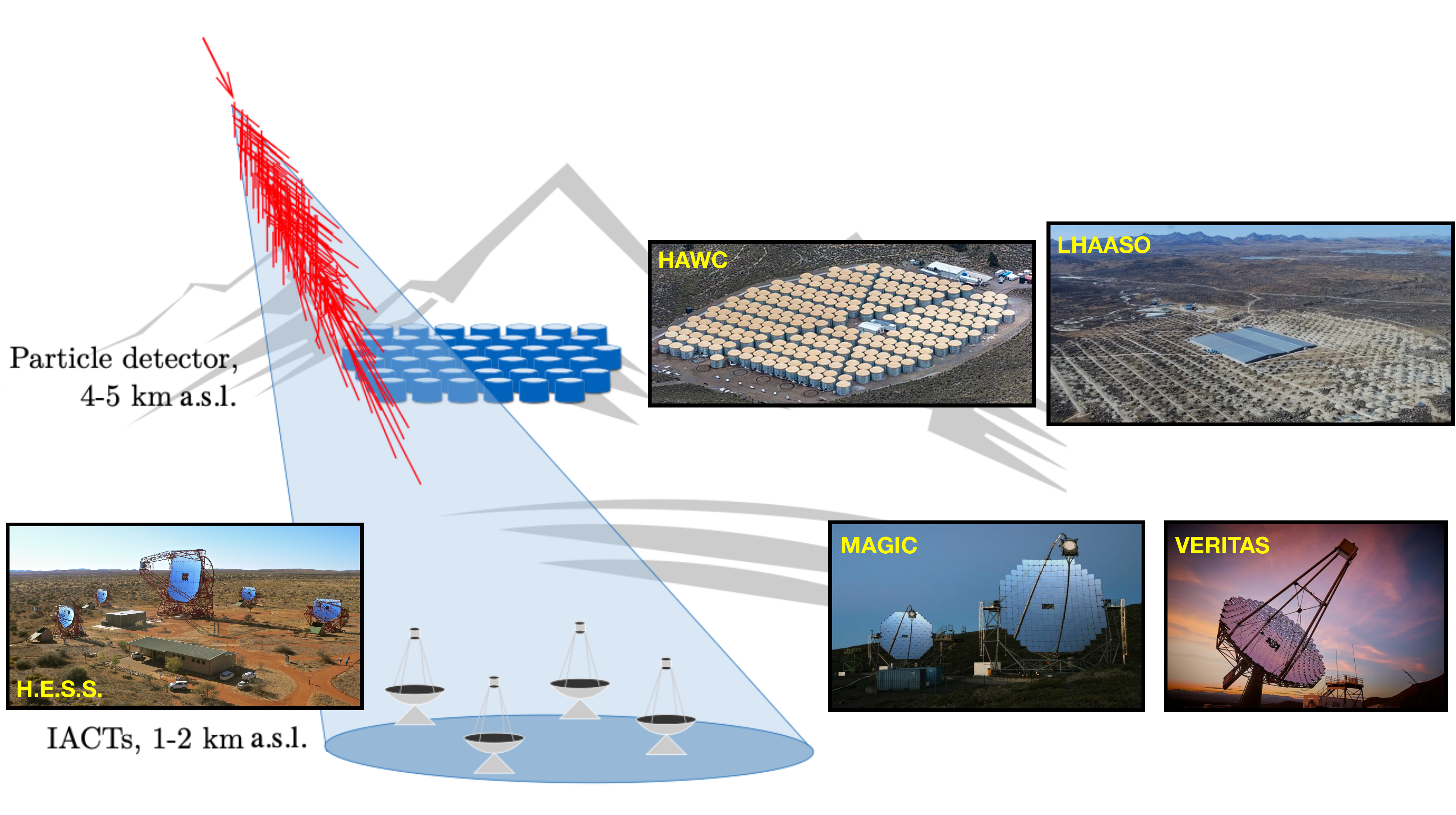}    
  \caption{Representation of different techniques for the detection of $\gamma$ rays: Imaging Atmospheric Cherenkov Telescopes (H.E.S.S., MAGIC and VERITAS) for very-high-energy $\gamma$ rays (0.1$-$100 TeV) and Extensive Air Shower experiments (HAWC and LHAASO) for very-high-energy to ultra-high-energy $\gamma$ rays (0.1 TeV $-$ 100 PeV). Figure adapted from \cite{SWGO:2019}.}
  \label{fig:Instru}
\end{figure}

After the first detection of the Crab nebula by the pioneering telescope Whipple in 1989 \citep{Weekes:1989_Crab} and following experiments, the current generation of IACTs were built and allowed the detection of $\gamma$ rays in the very-high-energy range, from $\sim$ 50 GeV to $\sim$ 50 TeV. $\gamma$ rays when interacting with the atmosphere produce an electromagnetic shower, as illustrated in Figure~\ref{fig:Instru}. The resulting secondary particles emit Cherenkov radiation that is detected by IACTs and allows the reconstruction of the direction and energy of the primary $\gamma$ ray. Located in the Southern hemisphere, H.E.S.S. (Namibia) is the most suitable to observe the inner parts of the Galaxy, but together with MAGIC (Canary Islands) and VERITAS (Arizona, United States) in the Northern hemisphere, they provide a full coverage of the Milky Way at very high energy. Detailed morphological and spectral analyses provided an unprecedented view of the TeV $\gamma$-ray sky, and after more than twenty years of observations, we have learned among other things that:

\begin{itemize}

\item Supernova remnants can be indeed prominent TeV emitters. However, the nature of their $\gamma$-ray emission is often difficult to understand (dominantly hadronic or leptonic), and most of them imply a proton cutoff energy well below PeV energies. Even Cassiopeia A, one of the youngest supernova remnants in the Galaxy ($\sim $ 350 years) which shows hadronic $\gamma$-ray emission, exhibits a cutoff at $E_{\rm{max, p}} \sim 10$ TeV \citep{MAGIC:2017_CasA}. This dramatically compromised the paradigm of supernova remnants as the main source of Galactic CRs. 

\item Pulsar wind nebulae are the most numerous identified Galactic sources at very high energy, among the 78 sources detected in the H.E.S.S. Galactic Plane Survey \citep{HGPS2018}. This can be explained by the fact that their $\gamma$-ray emission lasts significantly longer ($t \sim$ 100 kyr) than the one from supernova remnants. Moreover, the multi-wavelength non-thermal emission (from the radio to the TeV bands) from these pulsar wind nebulae is well explained with leptonic scenarios, with $\gamma$-ray emission often peaking in the H.E.S.S. energy range.

\item Young massive stellar clusters such as the Cygnus Cocoon and Westerlund 1 have $\gamma$-ray emission extending up to tens of TeV without a cutoff (making them PeVatron candidates). These very-high-energy sources that seem to continuously inject particles into the interstellar medium (derived from morphological studies) have been suggested as the main source of Galactic CRs \citep{Aharonian:2019_MSC}.

\item Some binary systems are also TeV emitters such as the $\gamma$-ray binary Eta Carinae \citep{HESS:2020_EtaCar} or the recurrent nova RS Ophiuci \citep{MAGIC:Ophi2022, HESS:2022_RSOphiuci}. Although it has been shown that $\gamma$-ray emission from some binary systems could be hadronic, the maximal energy reached by particles is well below hundreds of TeV. Recently, microquasars joined the class of binary TeV sources with the first detection of the jets in the SS433 system by HAWC \citep{HAWC:2018_SS433} which have been then deeply studied by H.E.S.S \citep{HESS:2024_SS433}.

\item Surprisingly, almost half of the detected sources from the H.E.S.S. Galactic Plane Survey are not identified. This is either due to a lack of multi-wavelength counterparts or crowded regions in which multiple sources could produce $\gamma$-ray emission. Some of these sources are PeVatron candidates with $\gamma$-ray emission up to 30 TeV without any cutoff. The so-called "dark TeV sources" (those without multi-wavelength counterparts) could be ancient pulsar wind nebulae with low magnetic field preventing the detection of synchrotron emission \citep{Tanaka:2010}, or molecular clouds illuminated by CRs \citep{Gabici:2009}.
\end{itemize}

Twenty years of deep observations of the Galactic sky provided a large variety of very-high-energy $\gamma$-ray emitting sources. Still, none of them show compelling evidence for hadronic acceleration up to PeV energies, and the paradigm of supernova remnants as the main source of Galactic CRs was drastically questioned. However acceleration up to PeV energies might occur during a very short period of their lifetime ($\lesssim 100$ years), so it is possible that no known supernova remnant acts \textit{today} as an active PeVatron. Unless we would be able to detect $\gamma$-ray emission during the first stage of their evolution, demonstrating that supernova remnants act as PeVatrons may require indirect evidence of escaped PeV particles interacting with nearby molecular clouds.

One of the main limitations of current IACTs in revealing PeVatrons is their sensitivity drop around 30 TeV. The lack of statistics at these energies prevents to reveal the highest-energy part of the spectrum of some PeVatron candidates. Current Extensive Air Shower (EAS) experiments, who recently joined the efforts on the search for PeVatrons, allowed to go beyond the very-high-energy range and opened with unprecedented sensitivity a new window on the Galactic $\gamma$-ray sky.

\section{Entering the ultra-high-energy $\gamma$-ray domain}\label{sec:EAS}
The development of EAS experiments located at higher latitudes ($\sim 4000$ m above sea level) allowed to probe higher-energy $\gamma$ rays by detecting the shower front of secondary particles (Figure~\ref{fig:Instru}). Thanks to a wide field of view and a high duty cycle, water Cherenkov tank detectors as installed at the HAWC (Mexico) and LHAASO (China) sites, provide a better sensitivity above 10 TeV than current IACTs but at the expense of a lower angular resolution \citep[0.12$^{\circ}$ at 50 TeV for HAWC compared to $\sim$ 0.07$^{\circ}$ with current IACTs in their optimal energy range,][]{HAWC:2024_Crabperf}. The kilometer square array (KM2A), composed of electromagnetic and muon detectors, was also installed at the LHAASO site in order to probe $>$ 10 TeV $\gamma$ rays, with an angular resolution of 0.5$^{\circ}$ at 20 TeV and 0.2$^{\circ}$ at 100 TeV \citep{LHAASO:2021_Crab_KM2Aperf}.

Since EAS are more suitable to detect highly extended sources, HAWC was able to reveal a new TeV source class: the pulsar halos of Monogem and Geminga, produced by escaped leptons from the pulsar wind nebula that diffuse through the interstellar medium \citep{HAWC:2017_Geminga_Monogem}. With a total livetime of 1039 days, HAWC was then the first to reveal $\gamma$ rays above 56 TeV within the Galaxy, with the detection of 8 extended sources (in addition to the Crab nebula) including 3 of them with a curved emission extending up to 100 TeV \citep{HAWC:2020_Cat}. Soon after, LHAASO also detected photons above 100 TeV from 12 Galactic sources although without clearly identifying their origin except for the Crab pulsar wind nebula \citep{LHAASO:2021_12sources}. But the biggest breakthrough came from the first LHAASO catalog, which contains 90 sources in total including 32 new sources (without any known very-high-energy counterpart), 75 sources with $E > 45$ TeV and 43 sources with $E > 100$ TeV \citep{LHAASO:2024_FirstCat}. These unprecedented multiple detections of $\gamma$ rays above 100 TeV opened the new window of the so-called ultra-high-energy $\gamma$-ray domain (100 TeV $-$ 100 PeV). Among the LHAASO sources, 35 are associated with an energetic pulsar among which 22 are detected above 100 TeV indicating that pulsar wind nebulae or halos might be prominent TeV emitters. The majority of the highest-energy sources was unexpectedly located near well-known lepton accelerators and did not clearly pinpoint any standard candidates for proton acceleration up to PeV energies. Within the LHAASO cataloged sources, 7 are categorized as dark sources (without association with the GeV, pulsar and supernova remnant catalogs). Three of the main implications from these discoveries are summarized below.

\subsection{Leptonic PeVatrons}

A photon of 1.1 PeV was detected from the Crab pulsar wind nebula by LHAASO, and the very well-constrained broadband spectrum can be easily explained by accelerated electrons with a maximal energy of $E_{\rm{max,e}} = 2.15$~PeV in a magnetic field of 112~$\mu$G \citep{LHAASO:2021_Crab}, as shown in Figure~\ref{fig:Crab_G106} (left). This first definitely showed that leptons can efficiently produce $\gamma$ rays above 100 TeV so the detection of ultra-high-energy photons can no longer be considered as unambiguous evidence for proton acceleration. For this reason the term \textit{leptonic PeVatron} (or electron PeVatron) was introduced in the literature. A possible steepening between 60 and 500~TeV, followed by a potential hardening of the spectrum at 1~PeV, was interpreted as an indication for an additional hadronic component. However this spectral transition is not significant at this stage and more statistics are needed above 1~PeV to draw a firm conclusion.

Given the fact that half of the ultra-high-energy sources detected by LHAASO are spatially coincident with an energetic pulsar, pulsar wind nebulae or halos may significantly contribute to the highest-energy $\gamma$-ray sources in the Galactic sky. In this sense, more leptonic PeVatrons could be revealed in the future. We note that it is however surprising that most of the energetic pulsars are associated with a LHAASO source but not all of them. This tends to indicate that the emission from pulsar wind nebulae (or halos) is affected by other pulsar parameters or the surrounding environment, besides the energetics.

\subsection{Supernova remnants as former Pevatrons?} 

The paradigm of supernova remnants was already challenged with the observation by IACTs of particle cutoff energies well below the PeV range, even in the youngest objects. LHAASO reported $\gamma$-ray emission from the old supernova remnant W51C ($t \sim$ 30 kyr) for which hadronic emission was confirmed through the detection of the pion bump feature \citep{Jogler:2016_W51C}. However, the spectrum is very soft ($\Gamma = 2.55 \pm 0.01$) and the best-fit model gives a maximal proton energy of $E_{\rm{max, p}} \sim$ 400 TeV \citep{LHAASO:2024_W51C}. Because W51C is already very old, its low-energy cutoff does not indicate that supernova remnants cannot act as PeVatrons during a certain period of their lifetime and could be therefore \textit{former PeVatrons}.

\begin{figure}[th!]
 \centering
 \includegraphics[width=1\textwidth, clip]{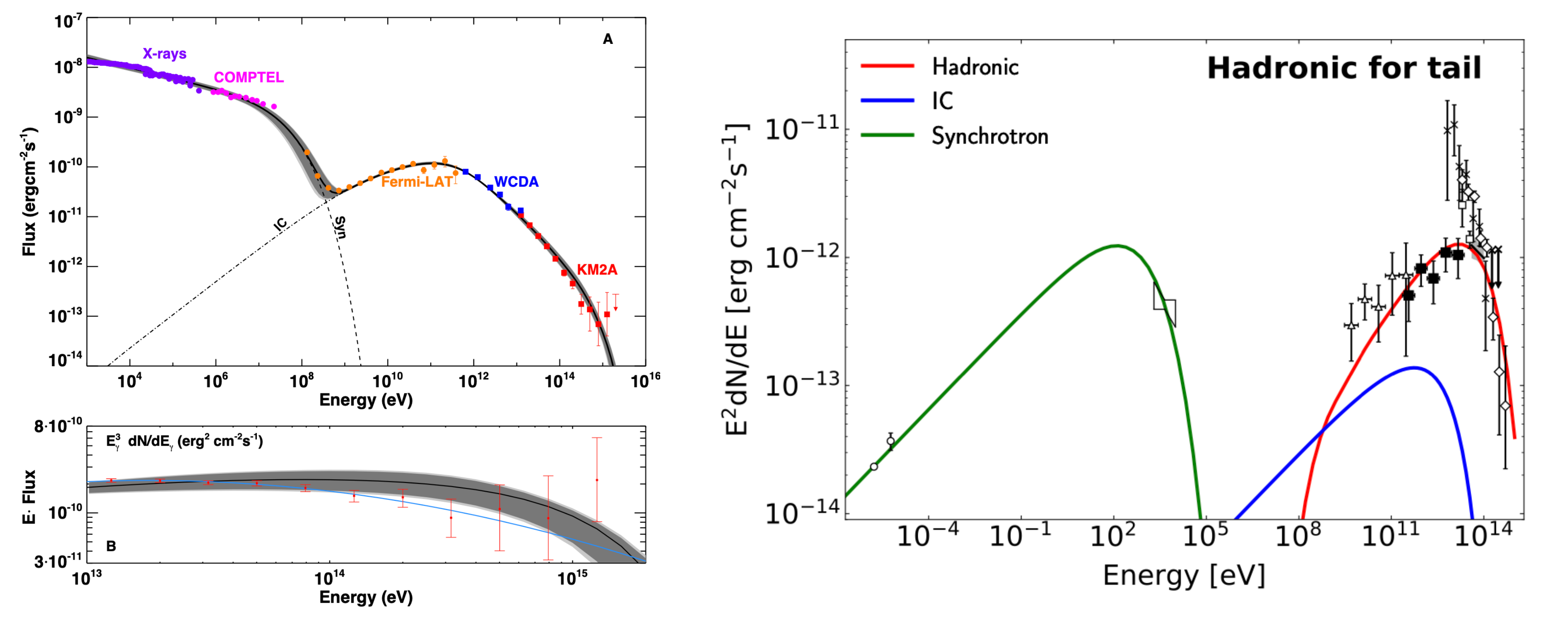}   
  \caption{\textbf{Left:} Best-fit leptonic model reproducing the broadband nonthermal spectrum of the Crab pulsar wind nebula (synchrotron and Inverse Compton emission). Figure extracted from \cite{LHAASO:2021_Crab}. \textbf{Right:} Best-fit model reproducing the radio, X-ray and $\gamma$-ray data towards the supernova remnant G106.3+2.7 (IC stands for Inverse Compton). LHAASO data points encompass the head and the tail of the very-high-energy emission while the emission from MAGIC (black filled dots) comes only from the tail and is explained by hadronic emission from escaped PeV CRs (red curve). Figure extracted from \cite{MAGIC:2023_G106}.}
  \label{fig:Crab_G106}
\end{figure}

This scenario was invoked to explain the ultra-high-energy $\gamma$-ray emission reported towards the supernova remnant G106.3+2.7 \citep{LHAASO:2021_12sources}, whose age is estimated to 10 kyr if it is associated to the nearby pulsar. MAGIC performed follow-up observations that revealed a hard spectrum spatially coincident with a molecular cloud \citep{MAGIC:2023_G106}. While a leptonic scenario cannot explain both the X-ray and ultra-high-energy $\gamma$-ray data, a second population of escaped PeV CRs interacting with a nearby molecular cloud can reproduce the broadband spectrum (Figure~\ref{fig:Crab_G106}, right). G106.3+2.7 could therefore be a former PeVatron. Unless we can derive the maximal energy reachable by particles in supernova remnants by observing their emission during their first stage of evolution ($t \lesssim 100$ years), addressing the question of whether these objects can act as PeVatrons might only be possible through indirect observations, i.e. detection of PeV particles outside their acceleration site. Future observations that could be explained only by this scenario would support the fact that supernova remnants are former PeVatrons.

\subsection{PeV protons towards the Cygnus Cocoon}

\begin{figure}[t!]
 \centering
 \includegraphics[width=1\textwidth,clip]{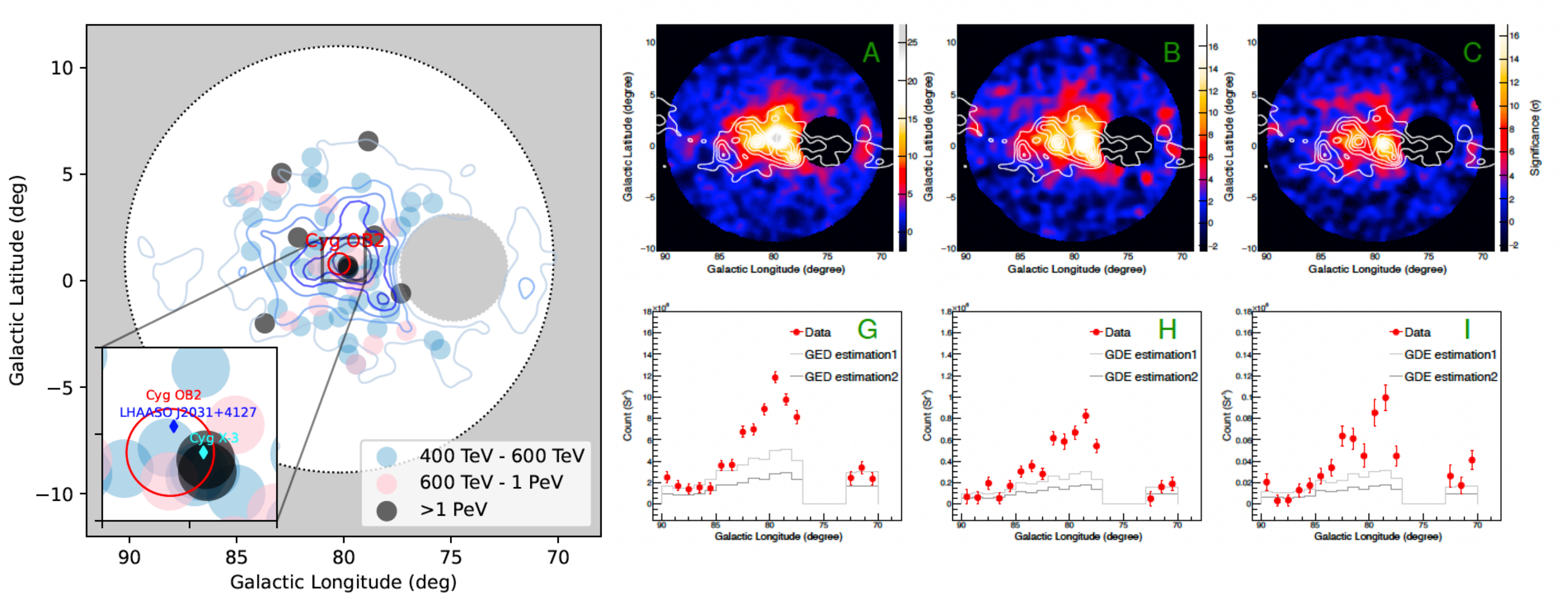}    
  \caption{\textbf{Left:} Events detected by LHAASO from the Cygnus Cocoon. The star forming region Cygnus OB2 is represented by the red circle. \textbf{Right:} Significance of the $\gamma$-ray emission in different energy bands (2$-$20~TeV, 25$-$100~TeV and $> 100$~TeV), after removing individual sources and with the gas overlaid in white. The corresponding longitude count profiles (red points) with two Galactic diffuse emission estimates (grey histograms) are also given. Figures extracted from \cite{LHAASO:2024_Cygnus}.}
  \label{fig:Cygnus}
\end{figure}

One of the most exciting results was the detection of PeV photons from the vicinity of the Cygnus Cocoon, a superbubble surrounding a region of OB2 massive star formation with a size of $\sim$ 150 pc \citep{LHAASO:2024_Cygnus}. The Cygnus Cocoon was already detected at very high energy \citep{AGRO:2012_Cygnus} showing a $\gamma$-ray emission that correlates with the gas and a lack of an energy-dependent morphology that supports the hadronic nature of the emission. LHAASO observations revealed multiple photons above 1 PeV in a 6$^{\circ}$ region, as illustrated in Figure~\ref{fig:Cygnus}. Although the spectrum is curved and best-fitted by a logarithmic parabola, it implies the presence of protons of at least 10 PeV. Potential accelerators could be the star forming region Cygnus OB2, the microquasar Cygnus X-3 or even related to the pulsar PSR J2032+4127. More statistics at the highest energy and better angular resolution are needed to unveil the morphology and spectrum in details and pinpoint the origin of the emission. 

It should be noted that the spectrum from the Cygnus Cocoon is curved, and up to now, HAWC and LHAASO did not reveal any hadronic $\gamma$-ray emission with an uncurved power law extending up to hundreds of TeV. The definition of hadronic PeVatrons has changed in the literature and now also applies to sources with a curved or broken hadronic $\gamma$-ray spectrum extending up to 100~TeV. It is therefore useful to introduce the term $\textit{CR knee PeVatrons}$ \citep{Auguner:2023} to define the sources accelerating the $\textit{bulk of hadrons}$ to at least PeV energies which are seen in the CR spectrum, i.e. CR accelerators without a break or curvature in their spectrum up to PeV energies or beyond.

\section{Conclusions and Outlook}

Two decades of observations with current IACTs gave us an unprecedented view of the very-high-energy Galactic $\gamma$-ray sky, highlighting the large diversity of $\gamma$-ray emitting sources such as supernova remnants, pulsar wind nebulae, pulsar halos, massive stellar clusters or binaries. The new window of the ultra-high-energy sky recently opened by HAWC and LHAASO revealed indeed the existence of multiple powerful particle accelerators. The first one to be identified is the Crab nebula, as a leptonic PeVatron, and surprisingly, half of the ultra-high-energy $\gamma$-ray sources are located close to an energetic pulsar, while supernova remnants can be at best assumed as former PeVatrons. The detection of multiple PeV photons in the direction of the Cygnus Cocoon might indicate that massive stellar clusters can act as PeVatrons, although the curvature of the spectrum still needs to be understood. Given these new results, the terms $\textit{leptonic PeVatrons}$, $\textit{hadronic PeVatrons}$ and $\textit{CR knee PeVatrons}$ are introduced in the literature.

The next generation of Cherenkov telescopes CTA will provide increased statistics above 30 TeV and better angular resolution than current IACTs. Located in both hemispheres (Chile and Canary Islands), CTA will give valuable insights into the very-high-energy $\gamma$-ray sky during the next decades. Fully operated since 2015 and 2021 respectively, HAWC and LHAASO are expected to monitor the Northern sky during the next 20 years. The SWGO (an experiment similar to HAWC and aimed to be installed in the Southern hemisphere) will be necessary to provide the same complementarity between IACTs/EAS (angular resolution versus sensitivity at ultra high energy) and to monitor the other parts of the Galactic plane.

The question whether pulsar wind nebulae are able to accelerate protons is still open, as is whether supernova remnants indeed act as PeVatrons during a certain amount of their lifetime. Several questions still hold concerning the contribution to the Galactic CRs from stellar clusters or possibly from other sources such as microquasars. It is not excluded that multiple components produce CRs at different energies, for example isolated supernova remnants being responsible for the low-energy part and massive stellar clusters sustaining the highest-energy flux of Galactic CRs. Still numerous questions need to be answered concerning particle acceleration in astrophysical sources, as those related to their escape and their propagation. Thanks to current and incoming facilities, the future is bright for Galactic $\gamma$-ray science and will certainly shed light on the century-old question of the origin of Galactic CRs.

\bibliographystyle{aa} 
\bibliography{Devin_S06}

\end{document}